\documentclass{ws-mpla}
\usepackage[super]{cite}
\usepackage{graphicx}
\begin{document}

\markboth{D. Bortoletto}
{The Importance of Silicon Detectors for the Discovery of the Higgs Boson and the Study of its Properties}

\catchline{}{}{}{}{}

\title{The Importance of Silicon Detectors for the Higgs Boson Discovery and the Study of its Properties}

\author{DANIELA BORTOLETTO\\
on behalf of the CMS Collaboration}

\address{Department of Physics, University of Oxford, Keble Road\\
Oxford, OX1 3RH, UK\\
daniela.bortoletto@physics.ox.ac.uk}

\maketitle

\pub{Received (Day Month Year)}{Revised (Day Month Year)}

\begin{abstract}

Recent studies are presented demonstrating the important role played by silicon detectors in the the discovery of the Higgs boson.
CMS is planning to replace its  in an extended  technical stop of the LHC in the winter of 2016 . We present results showing that this replacement 
will significant increase the sample of Higgs bosons that will be reconstructed enabling precision studies of this particle.

\end{abstract}

\ccode{PACS Nos.: include PACS Nos.}

\section{Introduction}	

In 2012 the ATLAS and CMS experiments announced to the world the discovery of a new boson \cite{HiggsDiscovery}. Both collaborations have continued to measure the properties of the new particle using the whole data sample delivered by the LHC during it first run (run1). In this run the CMS experiment collected  $pp$ collision data  corresponding to an integrated luminosity of about 5.1 fb$^{-1}$  and 19.7 fb$^{-1}$ at the center-of-mass energy of  $\sqrt{s} $ = 7 and 8 TeV respectively.  A similar amount of data was collected by ATLAS. The data have already allowed precision measurements 
of the Higgs Boson mass.  Spin-parity studies favour the scalar, J$^{P} $= 0$^{+}$ hypothesis \cite{SpinParity}. Furthermore the production and decay properties of the new particle appear to be consistent with those expected of the SM Higgs boson \cite{Couplings}.  In this review we focus on the importance of silicon sensors in this discovery and the need to replace them to maintain the performance needed to allow precision measurements in the challenging environment of the LHC hadron collider .

The LHC is currently undergoing a shutdown that began on 14 February 2013. Repairs are ongoing to consolidate the high-current splices and allow the machine to run 
very closely to the design energy of 14 TeV. In addition the luminosity will increase from the record of $7 \times 10^{33} cm^{-2} s^{-1}$  achieved in run1 to over $1 \times 10^{34} cm^{-2} s^{-1}$ .
 
In this paper we describe  the importance of silicon detectors in this discovery.  We first briefly discuss the CMS detector and especially its pixel detector. We then summarise the status of studies of the Higgs boson using two pairs of same-flavor, opposite-charge, well-identified and isolated leptons, $e^{+}e^{-}$ and $\mu^{+}\mu^{-}$, compatible $H\rightarrow ZZ$, where one or both the $Z$ bosons can be off-shell.  Finally, we discuss the challenges that the higher LHC luminosity  brings and describe the impact of the replacement of the current pixel detector on the Higgs physics program, focusing as an example on the $H\rightarrow ZZ \rightarrow 4\ell$ channel.

 \section{CMS}
 
CMS, which is described in detail in \cite{CMSexp}, is one of the two multipurpose experiments at the LHC.  It is 21.6 m long, 15 m in diameter, and weighs about 12,500 t. The  apparatus is characterised  by  a very large  superconducting solenoid of 6m internal diameter, providing a 3.8 T magnetic field.  The CMS inner tracker consists of silicon pixel and strip detectors.  The inner tracker,  lead tungstate crystal electromagnetic calorimeter (ECAL), and brass/scintillator  hadron calorimeter (HCAL) are all located inside the solenoid magnet. Muons are detected in gas-ionization detectors
embedded in the iron flux return placed outside the solenoid. Extensive forward calorimetry complements the coverage provided by the barrel and endcap detectors.  The experiment uses a coordinate system with the origin at the nominal interaction point,  the $x$ axis pointing to the center of the LHC ring, the $y$ axis pointing up with repeat to the LHC ring, and the  $z$ axis along the beam direction using a right-handed convention. The polar angle $\theta$  is measured from $z$ axis and the azimuthal angle $\phi$ is measured in the $x-y$ plane. The pseudorapidity is defined as $\eta = -\ln[\tan(\theta/2)]$.

The inner tracker measures charged particle trajectories within the range $|\eta| < 2.5$. It consists of 1440 silicon pixel modules arranged into 3 barrel layers and 2 endcap disks on each side. The silicon tracker includes 10 barrel layers and a total of 10 end caps disks instrumented with 15148 modules.  It provides an impact parameter resolution of 15 $\mu$m and a transverse momentum (p$_T$) resolution of about 1.5\% for 100 GeV particles\cite{CMStrk}. The ECAL contains 75848 lead tungstate crystals and provides coverage of  $|\eta| < 1.479$ in the  barrel region (EB), and $1.479 < |\eta|< 3.0$ in the two endcap regions (EE).   A preshower detector  consisting of two planes of silicon sensors interleaved 
 with a total of 3 radiation lengths of  lead is located in front of the EE. The ECAL energy resolution for electrons 
with transverse energy $E_T >45 $ GeV from the $Z  \rightarrow  e^+e^-$ decays is better than 2\% in the central region of the EB  ($|\eta | < 0.8$), and is between 2\% and 5\% elsewhere. For low-bremsstrahlung electrons that have 94\% or more of their energy contained within a 3 $\times$ 3 array of crystals, the energy resolution improves to 0.5\% for $|\eta|< 0.8$.\cite{CMScalo}   The HCAL is a sampling calorimeter with brass as the passive material and plastic scintillator  tiles serving as active material, providing coverage of $ |\eta | < 2.9$.  A hadron forward calorimeter  extends the coverage up to $ |\eta | < 5.2$. Muons are detected in the pseudorapidity range $ |\eta | < 2.4$, with detection planes made using three technologies: drift tubes, cathode strip chambers, and resistive-plate chambers. The global fit of the muon tracks matched to the tracks reconstructed in the silicon tracker results
 in an average transverse momentum resolution that varies between 1.8\% at $p_T$= 30 GeV to 2.3\% at $P_T$ = 50 GeV. \cite{CMSmuon}
 
\subsection{The CMS pixel detector and its impact on tracking}

With about 66  million $100 \mu m \times 150 \mu m$ detector elements, the pixel system is a crucial component of the all-silicon CMS inner tracker \cite{CMSpixel}, playing a critical role in tracking, vertex reconstruction, and the High Level Trigger (HLT).  It was designed to record with high efficiency and precision the first three space-points near the interaction region in operating conditions up to the nominal instantaneous luminosity  of $1 \times 10^{34} cm^{-2} s^{-1}$ and 25 ns colliding bunch spacing. Under these conditions, an average of about 25 simultaneous overlapping events, or pile up, are expected per bunch crossing. Track reconstruction in such a high-occupancy environment is challenging since it is difficult to obtain a high track-finding efficiency, while keeping the fraction of fake tracks small. Fake tracks are tracks that may be formed from a combination of unrelated hits or from a particle trajectory that is badly reconstructed through the inclusion of spurious hits.  The detector performance during run 1 of the LHC was extremely robust and it was able to handle the record luminosity of  $7 \times 10^{33} cm^{-2} s^{-1}$  and 50 ns bunch crossing that lead to a higher than expected  average pile up of 37.

The pixel detector consists of cylindrical barrel layers at radii of 4.4, 7.3 and 10.2 cm, and two pairs of endcap disks at z = 34.5 and 46.5 cm. 
Its 1440 modules  cover an area of about 1 m$^2$ and have 66 million pixels. The pixel system performs zero-suppression in the readout chips of the sensors,  through a pixel threshold corresponding to an equivalent charge of about 3200 electrons. Offline, pixel clusters are formed 
from adjacent pixels and must have a minimum charge equivalent to 4000 electrons while a a minimum ionizing particle deposits usually around 21000 
electrons.  A fast hit reconstruction algorithm is used during track seeding and pattern recognition. A more precise algorithm, based on cluster shapes templates that can take into account the deterioration due to radiation exposure\cite{PIXELAV} is used in the final track fit.

\begin{figure}[ph]
\centerline{
\includegraphics[width=2.5in]{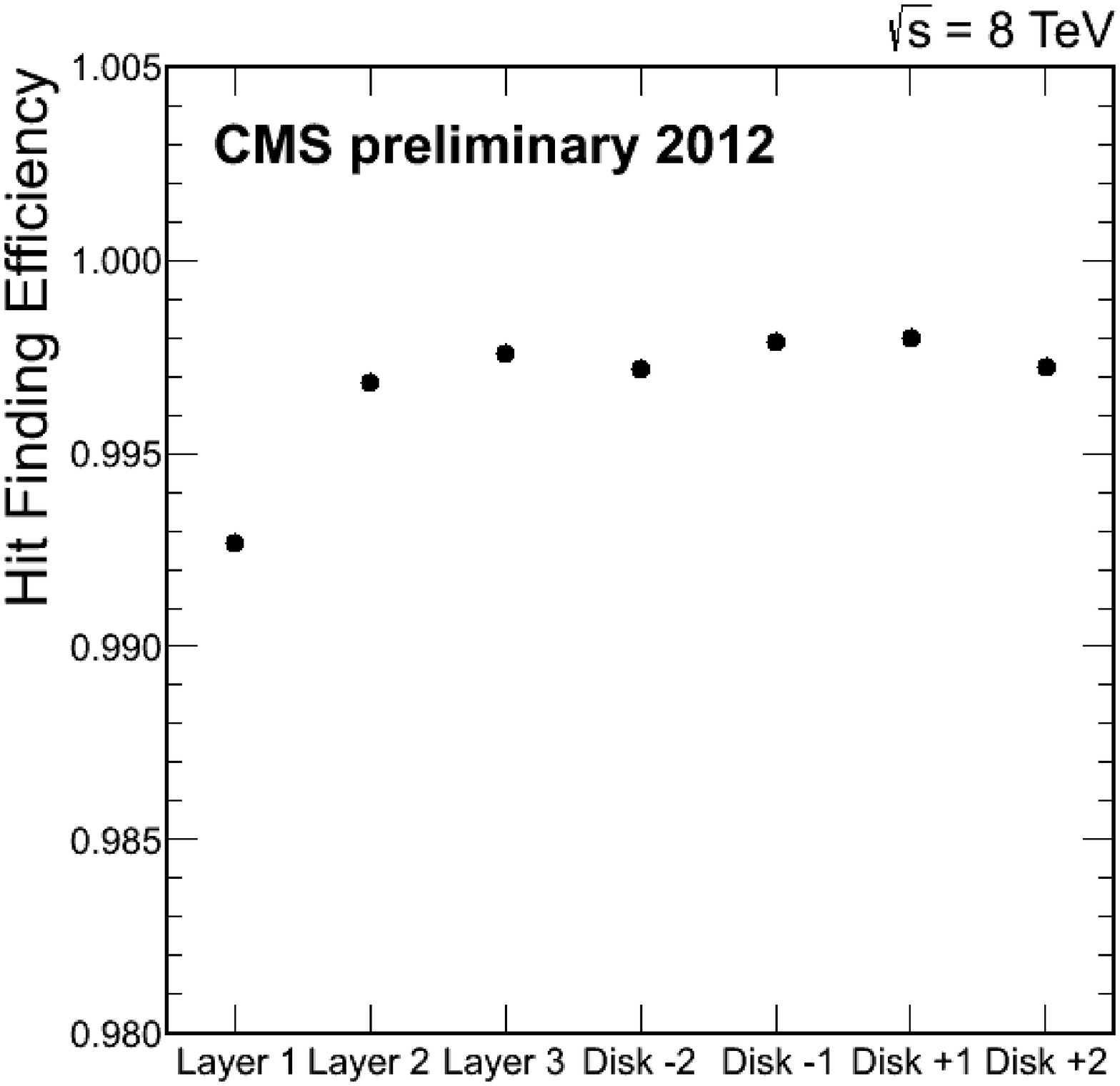}
\hspace*{8pt}
\includegraphics[width=2.5in]{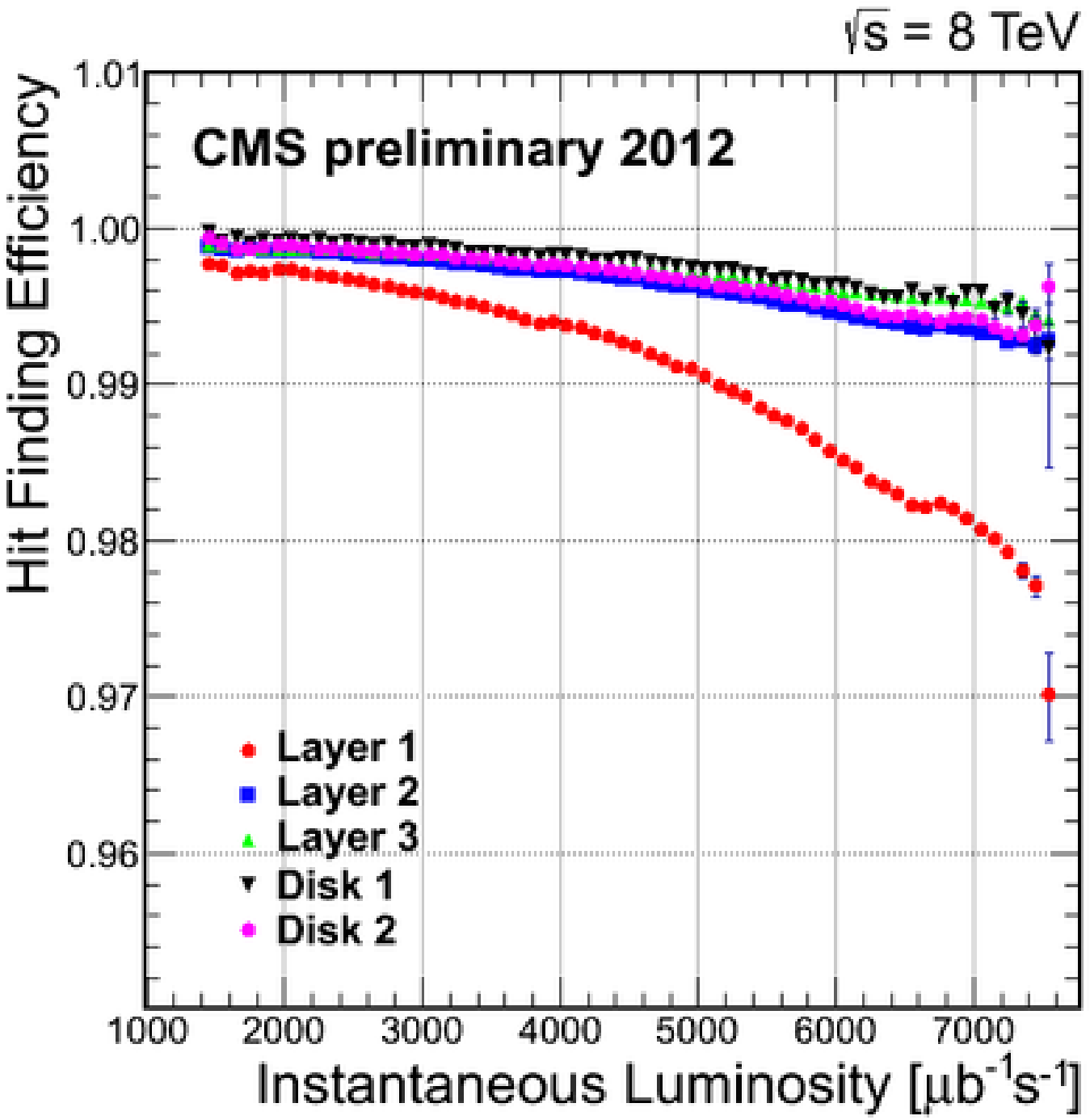}
}
\vspace*{8pt}
\caption{The average hit efficiency for layers or disks in the pixel detector excluding defective
modules (left), and the average hit efficiency as a function of instantaneous luminosity (right).
The peak luminosity ranged from 1 to 7 nb$^{-1}$s$^{-1}$ during the data taking.\protect\label{fig1}}
\end{figure}

The pixel efficiency is determined from the fraction of tracks which has a hit found within 500 $\mu$m of the predicted position of the track in the layer under study. The efficiency measured using isolated tracks originating from the primary vertex with $p_T> 1$ GeV is $ >99$\%, as shown in Fig.~\ref{fig1}.   In this measurement we have excluded the 2.4\% of the pixel modules known to be defective. We have also required the tracks to be reconstructed with a minimum of 11 hits  in the strip detector and  to have no additional trajectories within 5 mm  to reduce the effects due to high track density.  The module hit efficiency has been found to decrease as function of the instantaneous luminosity, as shown in the right plot of Fig.~\ref{fig1}. This is caused by a dynamic inefficiency which increases with hit rate and originates from limitations in the buffering of the readout chip. Therefore  the pixel modules in the innermost layer of the BPIX, layer 1 are the most affected.

The resolution of the pixel detector is measured  in the middle barrel layer by comparing the predicted position with the measured hit for tracks with $p_T > 12$ GeV, which are not affect by multiple scattering in the other layers . The resolution of the transverse coordinate is  around 9.4 $\mu$m. The resolution in the longitudinal, which depends on the angle of the track relative to the sensor and  the amount of charge sharing, varies between 29 and 40 $\mu$m. 

The tracking software at CMS, referred to as the Combinatorial Track Finder (CTF),  is based on the Kalman filter and allows both pattern recognition and track fitting.  The seed generation starts in the pixel detectors which have a low fraction of hits per channel (0.002 - 0.02\%) due to their high granularity. In addition the pixel detector offer a  3-D space point facilitating track finding and reconstruction. Once the seeds are selected,  the CFT starts estimating the track parameters provided by the trajectory seed, and then builds track candidates by adding hits from successive detector layers,  and updating the fit parameters at each layer. The CFT uses the location and uncertainty of the detected hits. In addition the amount of material crossed  allows to estimate the effects of multiple Coulomb scattering and energy loss.  Finally, the track trajectory is refitted using a Kalman filter from the inside outwards. The average track-reconstruction efficiency is 94\% in the barrel region ($|\eta|<0.9$) of the tracker and 85\% at higher pseudorapidity. Most of the inefficiency is caused by hadrons undergoing nuclear interactions in the tracker material. The track fake rate is  reduced by requiring a minimum number of layers that have 
hits, a good $\chi^2$/degrees of freedom in the fit, and compatibility with the primary interaction vertex.

The primary-vertex reconstruction determines the location of all proton-proton interaction vertices in an event. The resolution in jet-enriched  samples approaches 10 $\mu$m in $x$ and 12 $\mu$m in $z$ for primary vertices using at least 50 tracks.

CMS has also  developed tracks and primary vertices reconstruction based solely on pixel hits which is used for for the High Level Trigger (HLT) that  processes events at rates of up to 100 kHz. The pixel only algorithms are extremely fast, because they use only three layers and they take advantage of the low occupancy and high 3-D spatial resolution of the pixel detector.  The maximum efficiency for the pixel tracks is 85\%.
 
\section{The $H\rightarrow$ 4 leptons studies }

\begin{figure}[ph]
\centerline{
\includegraphics[width=2.5in]{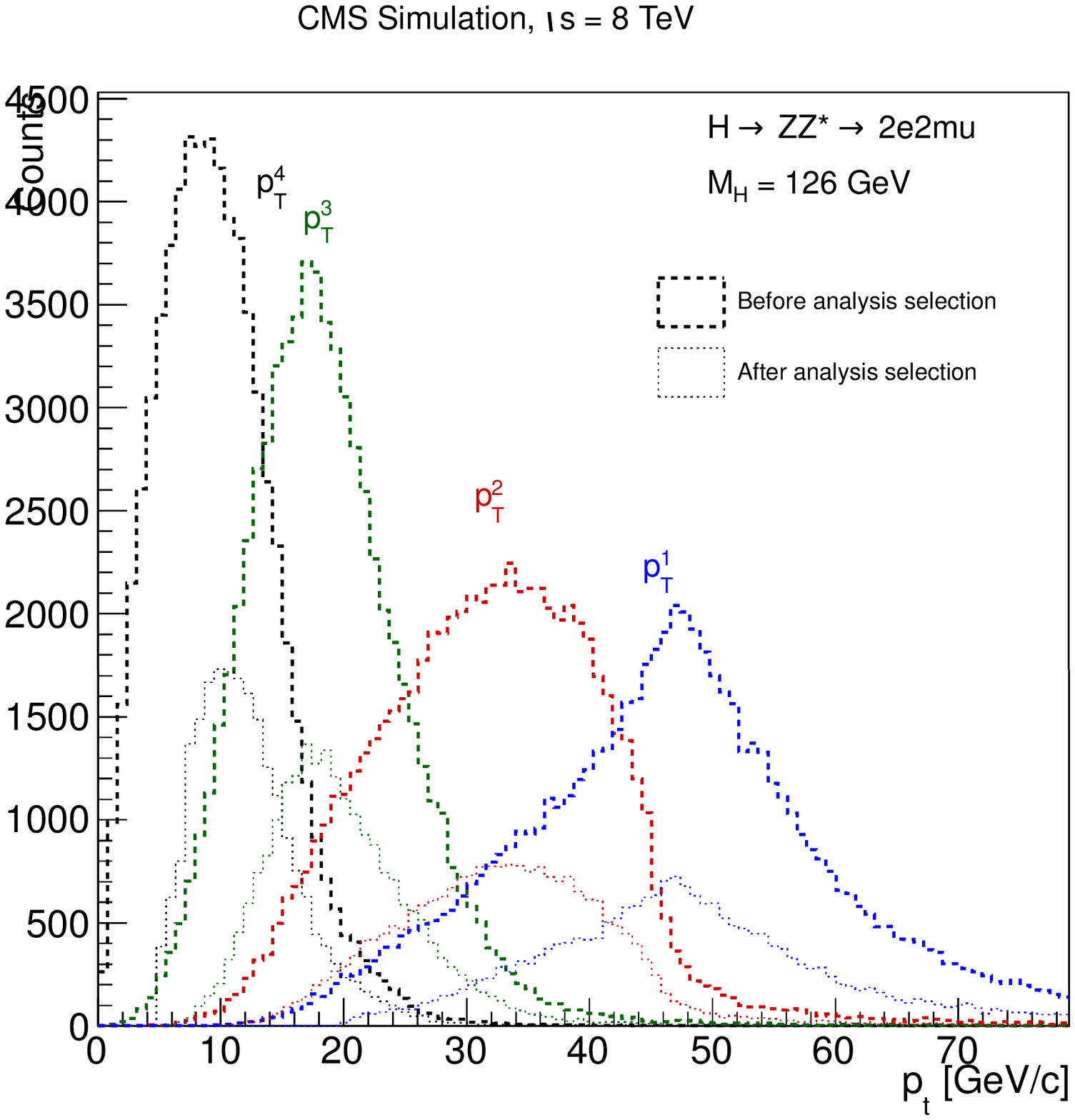}
\hspace*{8pt}
\includegraphics[width=2.5in]{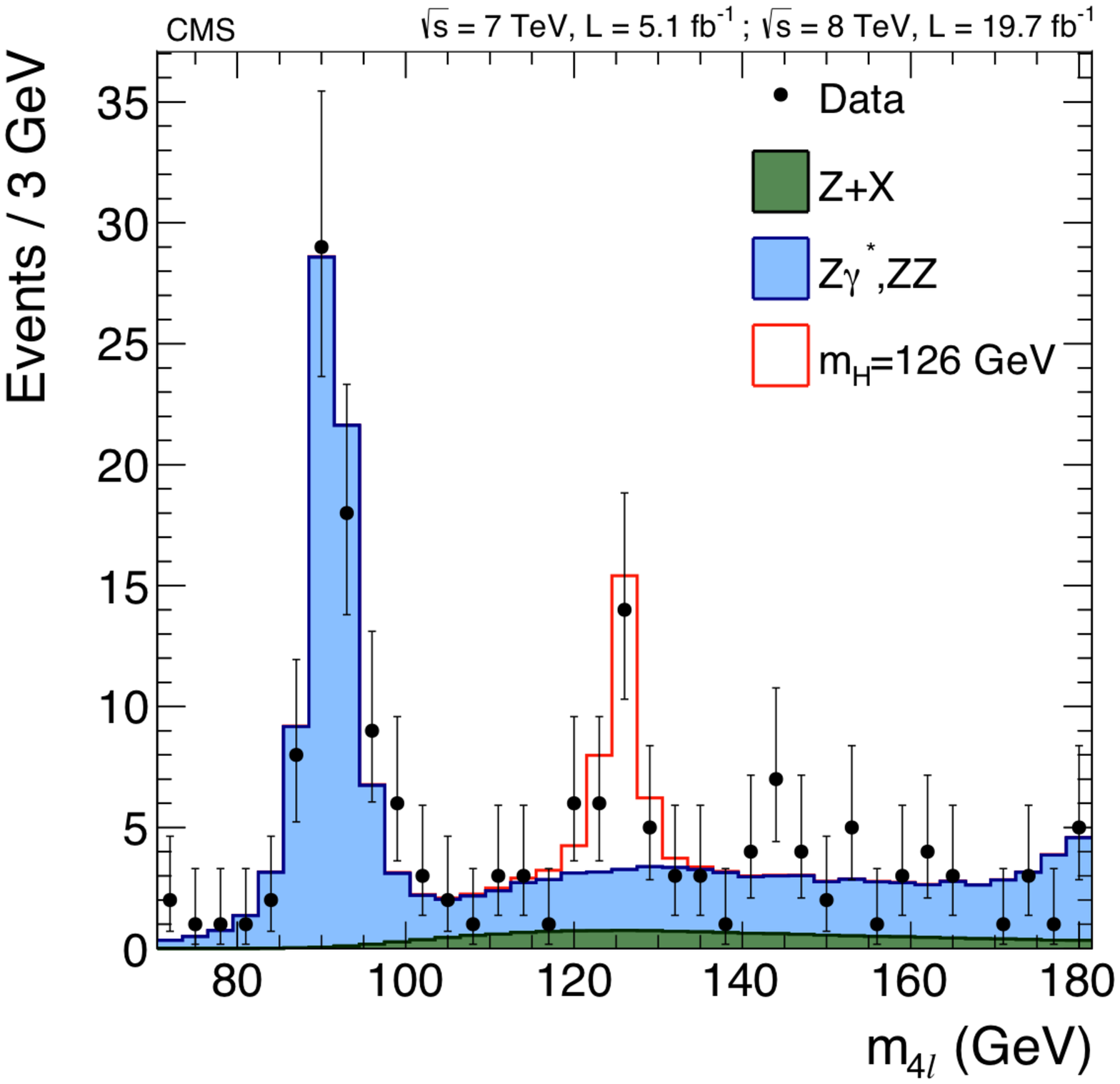}}
\vspace*{8pt}
\caption{Left: Momentum distribution of the 4 leptons in the $H\rightarrow ZZ\rightarrow 4 \mu$ final state
Right: Distribution of the four-lepton reconstructed mass for the sum of the $4e, 2e2\mu$, and $4\mu$ channels. 
Points with error bars represent the data, shaded histograms represent the backgrounds, and the unshaded histogram represents
the signal expectation for a mass hypothesis of $m_H = 126$ GeV. Signal and the $ZZ$ background
are normalized to the SM expectation; the $Z + X$ background to the estimation from data.g.\protect\label{fig2}}
\end{figure}

$H\rightarrow ZZ\rightarrow 4 \ell$  was the ÒgoldenÓ channel for the Higgs discovery in 2012\cite{HiggsDiscovery}. Since this final state is fully reconstructed and well measured, it continues to be critical  to the detailed study of the properties of the Higgs boson\cite{CMS4l}.  The branching fraction of the $H\rightarrow ZZ\rightarrow 4 \ell$ is  of O(10$^{-4}$) for $m_H$ = 125 GeV. Therefore  it is important to maintain a high lepton  selection efficiency over a range of 
momenta, to maximise the sensitivity for a Higgs boson of 125 GeV. In particular identification of  low $p_T$ electrons and muon is critical since the lowest momentum lepton in this decay has  $p_T$ well below 20 GeV, as shown in the left plot of  Fig.~\ref{fig2} for the $4\mu$ final state.  CMS has been able to identify muons and electrons with momentum above 7 and 5 GeV respectively. In addition since the signal appears as a narrow resonance on top of a smooth background  it is important to reach the best possible four-lepton mass resolution.  Several techniques have been deployed to calibrate the lepton momentum scale and resolution to a level
 such that the systematic uncertainty in the measured value of $m_H$  is far below the statistical uncertainty.  The CMS particle flow(PF) algorithm\cite{CMSPF}, which combines information from all sub-detectors,  is used to reconstruct  jets, missing transverse energy, and to determine lepton isolation observables.

The electron identification is optimised using a multivariate discriminant that combines observables sensitive to the bremsstrahlung along the electron trajectory, the momentum-energy matching between the electron trajectory and the associated supercluster, as well as ECAL-shape observables. The selection is  divided in six regions of the electron  $p_T$ and 
$ |\eta|$  due to the different performance. The calibration of the energy response is performed using a regression technique trained on a simulated dielectron  sample with the pileup conditions  matching the ones measured on data. This improves the resolution of the reconstructed invariant mass by 30\% for $H\rightarrow ZZ\rightarrow 4e$. The precision of 
  the electron momentum measurement is dominated by the ECAL and the tracker at high energies  and low p$_T$ respectively. The magnitude of the electron momentum is  determined by combining the two estimates with a multivariate regression function that takes as input the corrected ECAL energy from the supercluster regression, the track momentum estimate, their respective uncertainties, the ratio of the corrected ECAL energy over the track momentum as obtained from the track fit, the uncertainty in this ratio, and the electron category, based on the amount of bremsstrahlung. The absolute electron  momentum scale is calibrated using known resonances in data such as  the $J/\psi$, $\Upsilon(nS)$, and $Z$. The relative momentum scale between data and simulation is consistent within 0.2\% in the central barrel and up to 0.3\% in the end cap.

Muon candidates are required to have a transverse momentum $p_T > 5$ GeV and $|\eta| < 2.4$.  The reconstruction of the so called global muons combines information from both the silicon tracker and the muon system. To maximise the acceptance of low $p_T$ muons that may not penetrate the entire muon system tracker muons are accepted. These are tracks are tracks with matching hits in the muon chambers. The $p_T$ resolution for muons varies between 1.3 to 2.0\% in the barrel, and up to 6\% in the endcaps due to  multiple scattering of muons in the tracker material. The momentum scale and resolution for muons is studied using different resonances over different $p_T$ ranges. The agreement between the observed and simulated mass scales is within 0.1\% for $|\eta| < 2.4$.

Lepton isolation is required to discriminate leptons from the Higgs boson decay from leptons produced in hadronic processes which are usually part of jets. Leptons from in-flight decays of hadrons and muons from cosmic rays are suppressed by requiring that leptons are originating from the same primary vertex. 

At least one of the four-lepton candidates must have $p_T > 20$ GeV and another one is required to have $p_T > 10$ GeV.
The  first Z, denoted as $Z_1$, is selected by choosing the opposite-charge lepton pairs in the event with an invariant mass  closer to the $Z$ mass 
and in the range $40 < m_{Z_1}< 120$ GeV. Photons from final state radiation (FSR) are taken into account forming the Z-boson candidate if their inclusion makes the lepton pair mass closer to the nominal $Z$-boson mass. The remaining leptons are considered for the second  $Z$ pair, $Z_2$, if their invariant mass is $12< m_{Z_2}< 120$ GeV. We also require the four selected leptons to have $m_{\ell_1\ell_2} > 4$ GeV to reduce the contribution of lepton from hadronic decays. The  detection efficiencies, including geometrical acceptance, for a  Higgs boson with $m_H$ = 126 GeV are about  20\%, 29\%, and 42\%  for the $4e, 2e2\mu,$ and $4\mu$ channels respectively.  The resolution 
of the Gaussian core of the mass distribution, estimated from simulated signal samples with a double-sided Crystal-Ball function fit, is
about 2.0, 1.6, 1.2 GeV for the  $4e, 2e2\mu$, and $4\mu$  respectively. The uncertainty in the four-lepton mass is  estimated on a per-event basis using sum in quadrature of the individual mass uncertainty contributions from each lepton and any identified FSR photon candidate due to their momentum uncertainty. 

The production of Higgs boson through the Vector Boson Fusion mechanism can be be distinguished from  gluon fusion, which is about one order of magnitude larger, by requiring two high momentum jets at high $\eta$ and with large large separation $\Delta \eta$.  These two jets are the remnants of the incoming proton beams. Therefore the 4 lepton events are separated in events with zero or one jets and events with two jets. A linear discriminant which uses $\Delta \eta$ and the invariant mass of the two leading jets ($m_{jj}$) is used to distinguished gluon fusion from VBF events.

The expected yield and shape of the irreducible $ZZ$ background is evaluated by simulation using GG2ZZ\cite{GG2ZZ} for the $ gg \rightarrow ZZ$ contribution and POWHEG\cite{POWHEG} $q\overline{q }\rightarrow ZZ$.  The  Next-to-Leading Order (NLO) cross section for $q\overline{q }\rightarrow ZZ$ production and the Leading Order (LO) cross section for $ gg \rightarrow ZZ$  production are calculated with MCFM\cite{MCFM} .The reducible background due to $Z$ + jets, $t \overline{t}$, and $WZ$ + jets are determined from data using two different methods. One of the methods estimates the background using same-sign leptons, the other uses opposite sign leptons passing relaxed identification requirements when compared to those used in the analysis.

The reconstructed four-lepton invariant mass distribution for the combined $4e, 2e2\mu$, and $4\mu$ channels  shown in the right plot of Fig.~\ref{fig2}  displays a peak around 126 GeV over the expectations from background processes.

\begin{figure}[ph]
\centerline{
\includegraphics[width=3.75in]{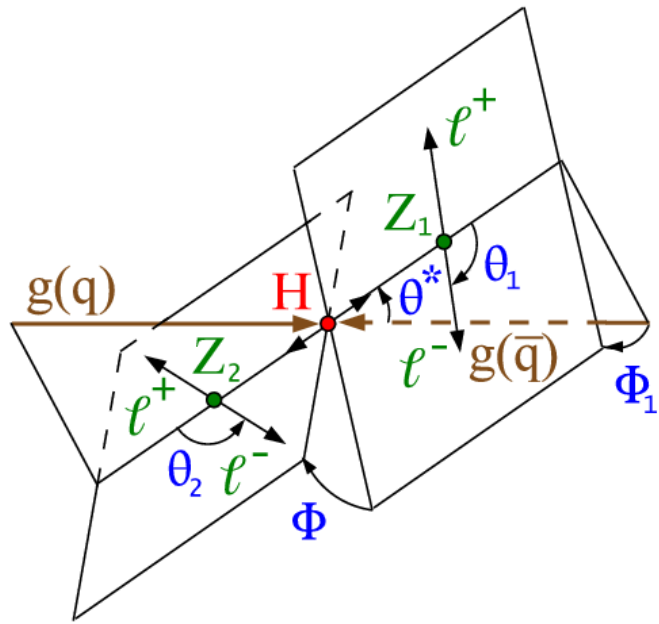}
\hspace*{8pt}
}
\caption{Illustration of the production and decay of a particle $H\rightarrow ZZ \rightarrow 4 \ell$
with the two production angles $\theta$ and $\phi$  shown in the H rest frame and three decay angles $\theta_1$, 
$\theta_2$, and $\Phi$ shown in the $Z_1$, $Z_2$, and $H$ rest frames, respectively.\protect\label{fig3}}
\end{figure}

The $4\ell$ final state is extremely powerful  since  the invariant masses of the lepton pairs, $m_{Z_1}$ and $m_{Z_2}$ and  all five angles ($\Omega=\Theta^*,\phi_1, \theta_1, \phi_2,\Phi$) defined in Fig.~\ref{fig3}  can be measured offering the possibility  to probe in detail the properties of the Higgs boson including its spin-parity.  Currently we do not have enough data to probe directly the tensor structure of the Higgs boson decay. Nonetheless we are able to build discriminants to compare the probability for the signal spin-parity hypothesis  ($0^+$) with respect to alternate $J^P$ hypothesis. 
The Higgs boson signal and background yields and the properties of the Higgs boson such as its mass and width, and the spin-parity quantum numbers, are determined with unbinned  maximum-likelihood fits performed to the selected events. The fits include probability density functions for five signal components (gluon fusion, VBF, WH, ZH and ttH production modes) and three background processes ($q\overline{q}\rightarrow ZZ$, $gg\rightarrow ZZ$ and $Z + X$).  The probability distribution of the of the angular and mass observables  ($\Omega ,m_{Z_1} ,m_{Z_2}$ ) and the  $m_{4\ell}$ distribution are used to enhance the signal to background discrimination. The  normalization of these components and systematic uncertainties are introduced in the fits as nuisance parameters, assuming log-normal a priori probability distributions, and are profiled during the minimization. The shapes of the probability density functions for the event observables are also varied within alternative ones, according to the effect induced by experimental or  theoretical systematic uncertainties. 

\begin{figure}[ph]
\centerline{
\includegraphics[width=2.5in]{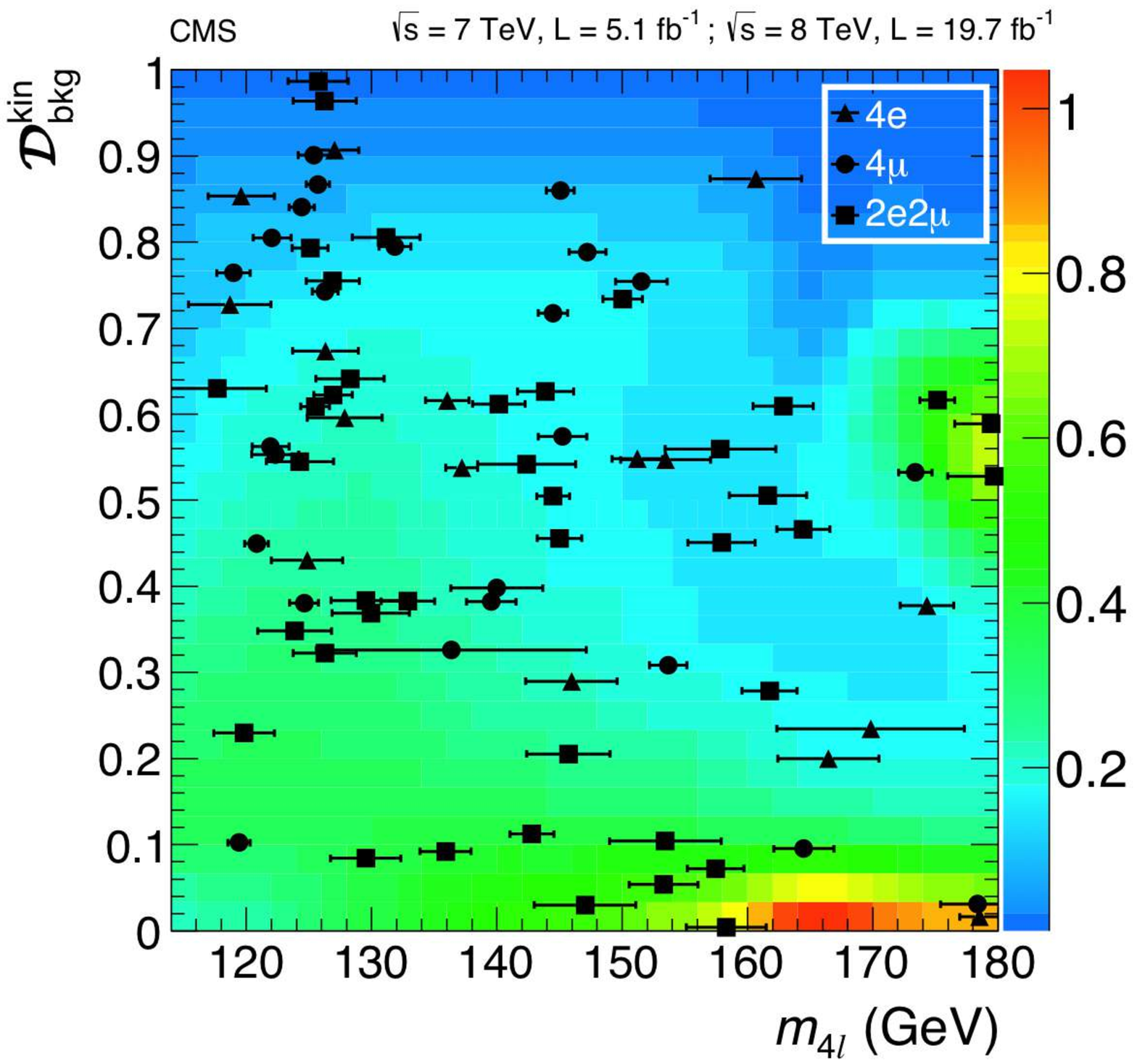}
\hspace*{8pt}
\includegraphics[width=2.5in]{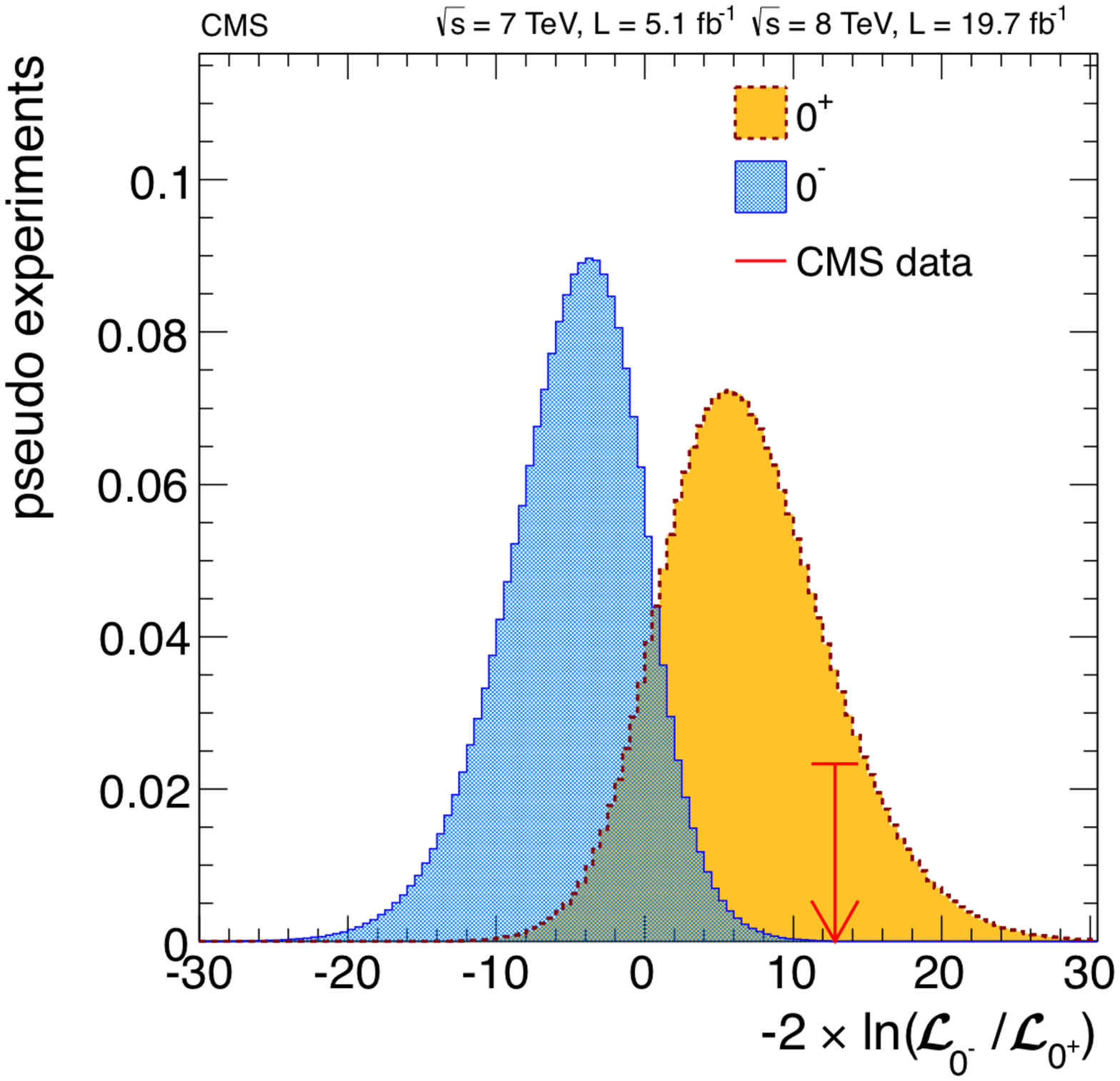}
}
\vspace*{8pt}
\caption{Left: Distribution of the kinematic discriminant $P(D^{kin}_{bkg}$ versus the four-lepton reconstructed
mass $m4\ell)$ The colour scale represents the expected relative density in linear scale (in arbitrary units) of background events. The points
show the data and the measured per-event invariant mass uncertainties as horizontal bars; Right: Distribution of the test statistic  for the SM pseudoscalar boson hypothesis (yellow) tested against the alternative $0^+$ hypothesis (blue). The arrow indicates the observed value\protect\label{fig4}}
\end{figure}

Several discriminant are built and optimised according to specific result to be extracted. Kinematic discriminants denoted  as $D^{kin}_{bkg}$ and 
$D^{JP}$ are defined based on the event probabilities depending on the background and the signal spin-parity ($J^P$) hypotheses under consideration respectively.The distribution of the kinematic discriminant  versus the four-lepton reconstructed mass $m_{4\ell}$ in the  low-mass region 
is shown in Fig.~\ref{fig4} . A signal-like clustering of events stands out at high values of  $D^{kin}_{bkg}$ and for $m_{4\ell}$ around 126GeV.
The signal significance, and the measurement of the Higgs boson signal strength, $\mu =\sigma/\sigma_{SM}$, defined as the measured cross section
times branching fraction into $ZZ$, relative to the expectation for the SM Higgs boson, is measured using 3D likelihood functions which include the 
$m_{4\ell}$, $D^{kin}_{bkg}$, and $p^{4\ell}_T$  separately for 0/1 jets  and 2 jets events. Using the full run 1 data set we find a local significance of 6.8$\sigma$, 
consistent with the expected sensitivity  of 6.7 $\sigma$ at a mass of $m_{4\ell}$ = 125.7 GeV.
The mass is measured using a 3D likelihood function with  the per-event mass uncertainty.  The mass is obtained through  profile scan of the 3D likelihood function versus the SM boson mass performed under  the assumption that its width is much smaller than the detector resolution. The combination of the 7 and 8 TeV data yields $m_H = 125.6 \pm 0.4 (stat.) \pm 0.2 (syst.)$GeV. The systematic  uncertainties account for the effect on the mass scale of the lepton momentum scale and resolution, shape systematics in the $P(D^{kin}_{bkg},m4\ell)$ probability density functions,  signal and background modelling, and normalization systematics due to acceptance and efficiency uncertainties.  Other measurements using 2D and 1 D likelihood scan provide consistent results.

The spin-parity hypothesis tests, are performed with 2D-likelihood functions based on $D_{bkg}$ and $D^{JP}$. Several alternative spin-parity model have been tested by forming the ratio of signal plus background  likelihoods for two signal hypotheses $q = -2ln(L_{J^P}/L_{0^+})$. The expected distribution of $q$ for the pseudoscalar hypothesis (blue histogram) and the SM Higgs boson (orange histogram)  are shown in Fig. ~\ref{fig4}. Similar distributions for the test statistic $q$ are obtained for  other alternative hypotheses. The pseudo-experiments are generated using the nuisance parameters fitted in data. The pseudoscalar ($0^-$) and all spin-one hypotheses tested are excluded at the 99.9\% or higher CL. All tested spin two models are excluded at the 95\% or higher CL. The $0^+$  hypothesis  is consistent with the data.

\section{The Phase1 pixel upgrade}

Three long shutdowns, designated  as LS1, LS2, and LS3 will be used to increase the luminosity of the LHC. In LS1, which is currently underway, the centre of mass energy 
will reach almost 14 TeV. In the period through LS2 (2018), the injector chain will be upgraded to deliver brighter bunches  into the LHC. In LS3 ($\approx$ 2022), the LHC itself will be upgraded with new low-$\beta$ triplets and crab-cavities to optimise the bunch overlap at the interaction
region.  We expect  that the machine  will deliver $1\times 10^{34}cm^{-2}s^{-1}$  soon after LS1. 
Furthermore it is anticipated that the peak luminosity will be close to  $2\times 10^{34}cm^{-2}s^{-1}$ before LS2, and perhaps significantly
higher after LS2. Currently the machine plan is to operate at 25 ns after LS1, but further 50 ns operation is not excluded. 
Therefore  CMS must be prepared to operate for the rest of this decade with average pile up of 50 as a baseline.
The performance of the current pixel detector will degrade in these conditions. For example, the hit efficiency will worsen
due to limits in the internal readout chip buffers and single event upsets. At the nominal L1A accept rate of 100 kHz, the data loss will increase to 16\% in the
innermost layer as the luminosity goes up by a factor of two (for 25 ns bunch crossing) to $ 2 \times10^{34} cm^{-2} s^{1}$.

The tracking efficiency of the current pixel detector is excellent, but slowly degrades  as the number of pile up events increases 
until it reaches a pile up of 40 when it begins to rapidly degrade due to filling buffers on the readout chip. 
There is also a noticeable dip in the efficiency in the pseudorapidity region near $|\eta| = 1.5$ where the bulkhead with the BPIX services overlap with FPIX.
The impact of material is also noticeable in the deterioration of momentum resolution of  tracks  that pass through theBPIX cooling pipes.

CMS is planning to replace the current pixel detector in an extended technical stop between 2016 and 2017. The design specification of the new pixel detector\cite{NewPixel}, denoted as Phase 1 pixel detector,  is that it should achieve
at $ 2 \times10^{34} cm^{-2} s^{1}$  the same or better performance that the current pixel detector  reaches at low luminosities of Run1. The upgrade detector
will  reuse  some of the current infrastructure and adopt DC-DC power converters and higher bandwidth electronics. The production and maintenance of the Phase 1 detector will be simpler since both barrel and forward pixel use only a module type with 16 readout chip organised in a $2 \times 8$ layout.  A new smaller diameter beam pipe has been recently installed 
to allow the placement of the inner pixel layer closer to the interaction region. A new pixel readout chip (ROC)\cite{NewRoc}  has been designed to minimize data loss due to latencies and limited buffering in high luminosity running. The column drain speed and the buffer depth for both the data (hits) and time stamp attached to 
the data will be increased. The size of the buffers has been optimised with detailed data flow simulations in the ROC. The new ROC has 24 time stamp buffer 
cells (12 for the present ROC) and 80 data buffer cells (32 for the present ROC). The data loss due to
buffer overflows at fluences up to 600 MPix/sec/cm$^2$ is less than 0.5\%. In addition the digital readout removes the need for the complex decoding of a multilevel analog signal.
The power distribution system in the ROC has changed substantially and several  signals have been rerouted and better shielded. These changes lead to a decrease
in absolute threshold  from the current 2800 down to 2000 electrons.  The new ROC has already undergone a couple of submission and the preliminary testing results are very positive.

\begin{figure}[ph]
\centerline{
\includegraphics[width=2.5in]{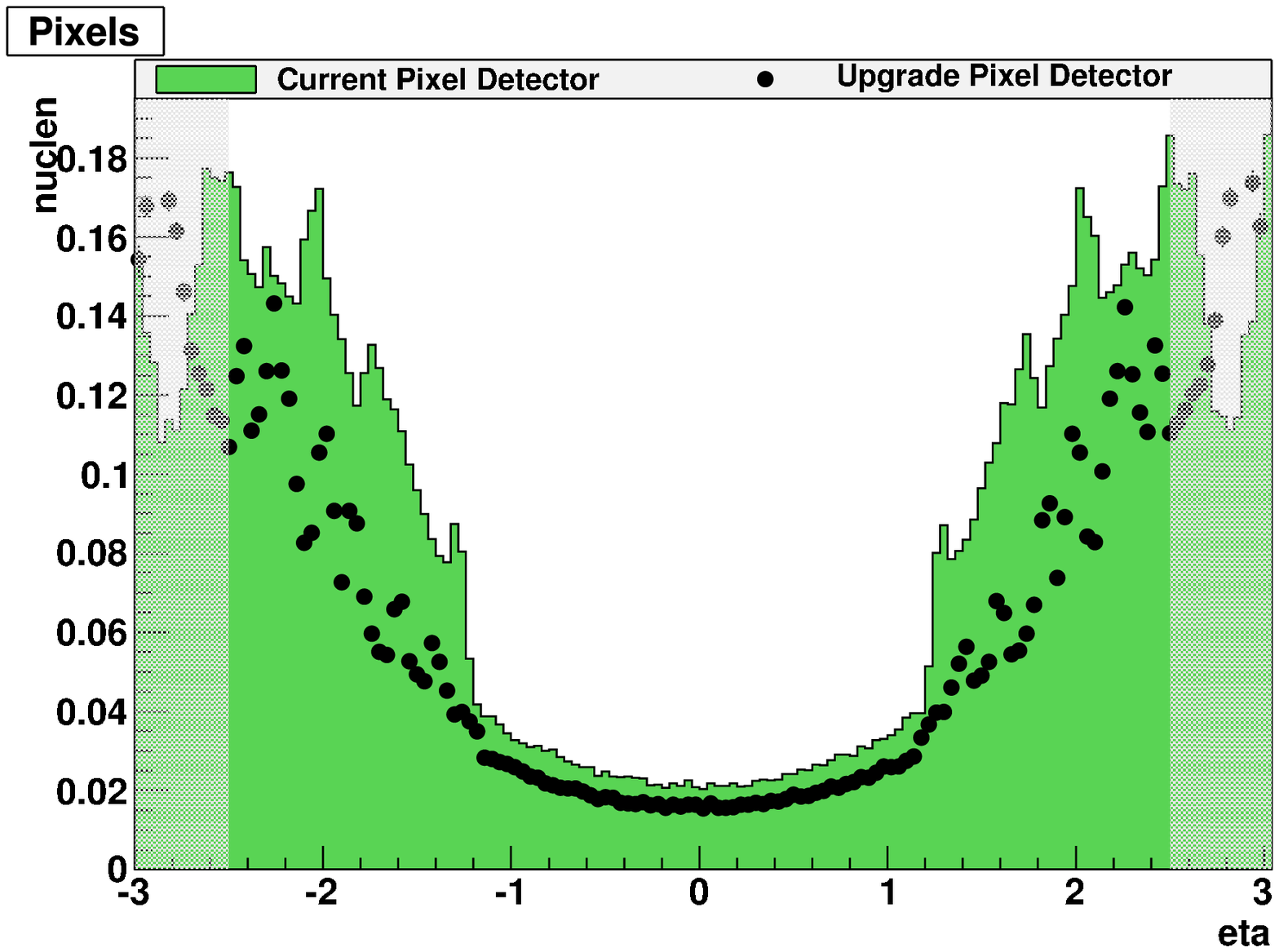}
\hspace*{8pt}
\includegraphics[width=2.5in]{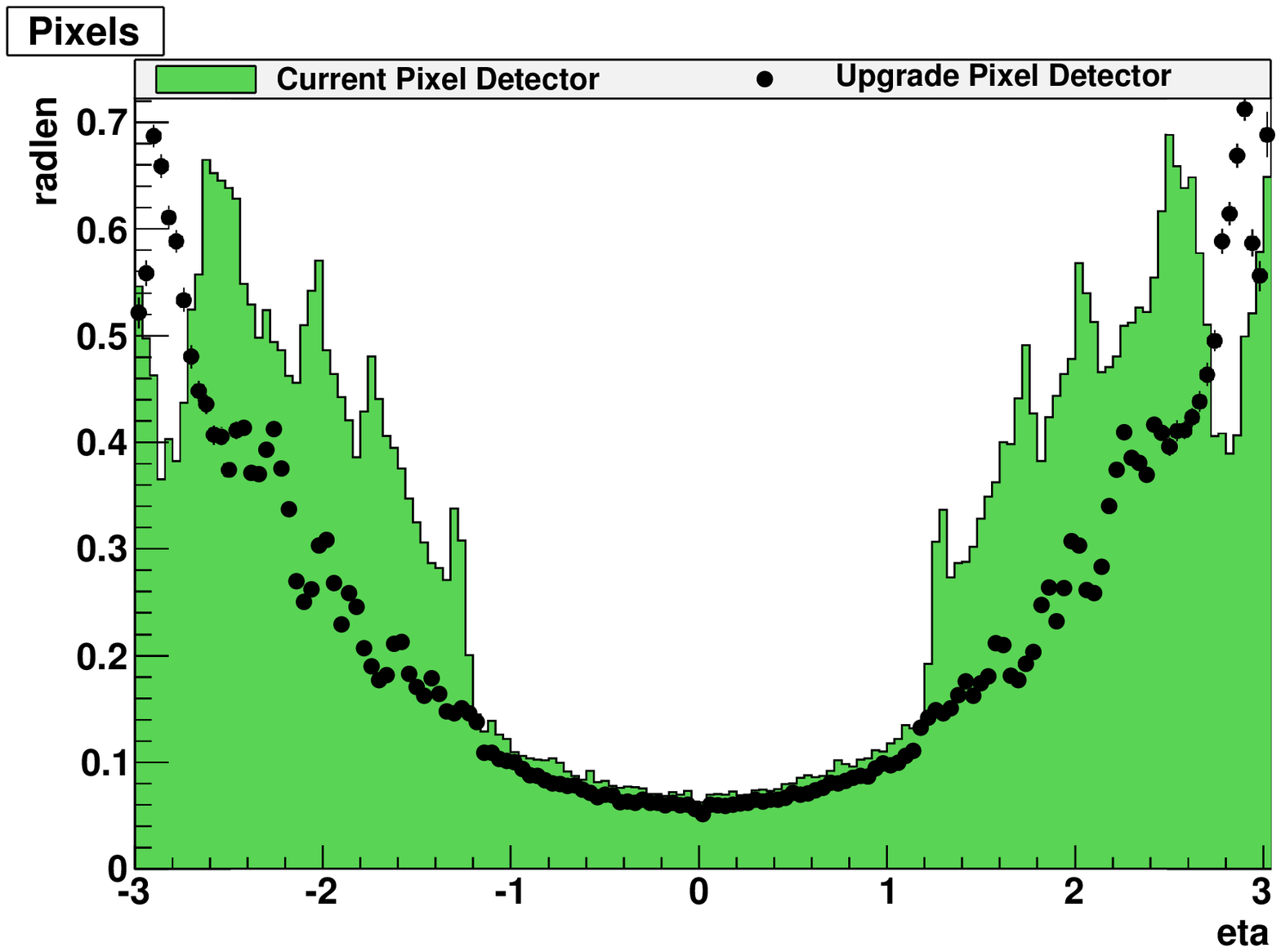}
}
\vspace*{8pt}
\caption{The amount of material in the pixel detector shown in units of radiation length (left), and in units of nuclear interaction length (right) as a function of h; this is given for the current pixel detector (green histogram), and the Phase 1 upgrade detector (black points). The shaded
region at high $\eta$ is outside the region for track reconstruction.\protect\label{fig5}}
\end{figure}

The Phase 1 pixel detector with four pixel-hit coverage over the $\eta$ range and a innermost layer at smaller radius
improves pattern recognition and track reconstruction. The addition of the fourth barrel layer at a radius of 16 cm also yields some safety margin in case the first silicon strip
layer of the Tracker Inner Barrel (TIB) degrades more rapidly than expected. It also provides the necessary  redundancy in pattern recognition and  contributes to reducing fake rates in
high pile up environment.The material in the tracking region has  been significantly reduced by
 adopting two-phase CO$_2$ cooling, light-weight mechanical support,  and by moving the electronic boards and connections out of the tracking volume. These improvements
 lead to a  reduction in the radiation length and nuclear interaction length of the present and upgrade pixel detectors as a function of $\eta$ as shown in Fig. ~\ref{fig5}.

\section{Impact of the Phase 1 pixel detector on Higgs physics}

\begin{figure}[ph]
\centerline{
\includegraphics[width=2.5in]{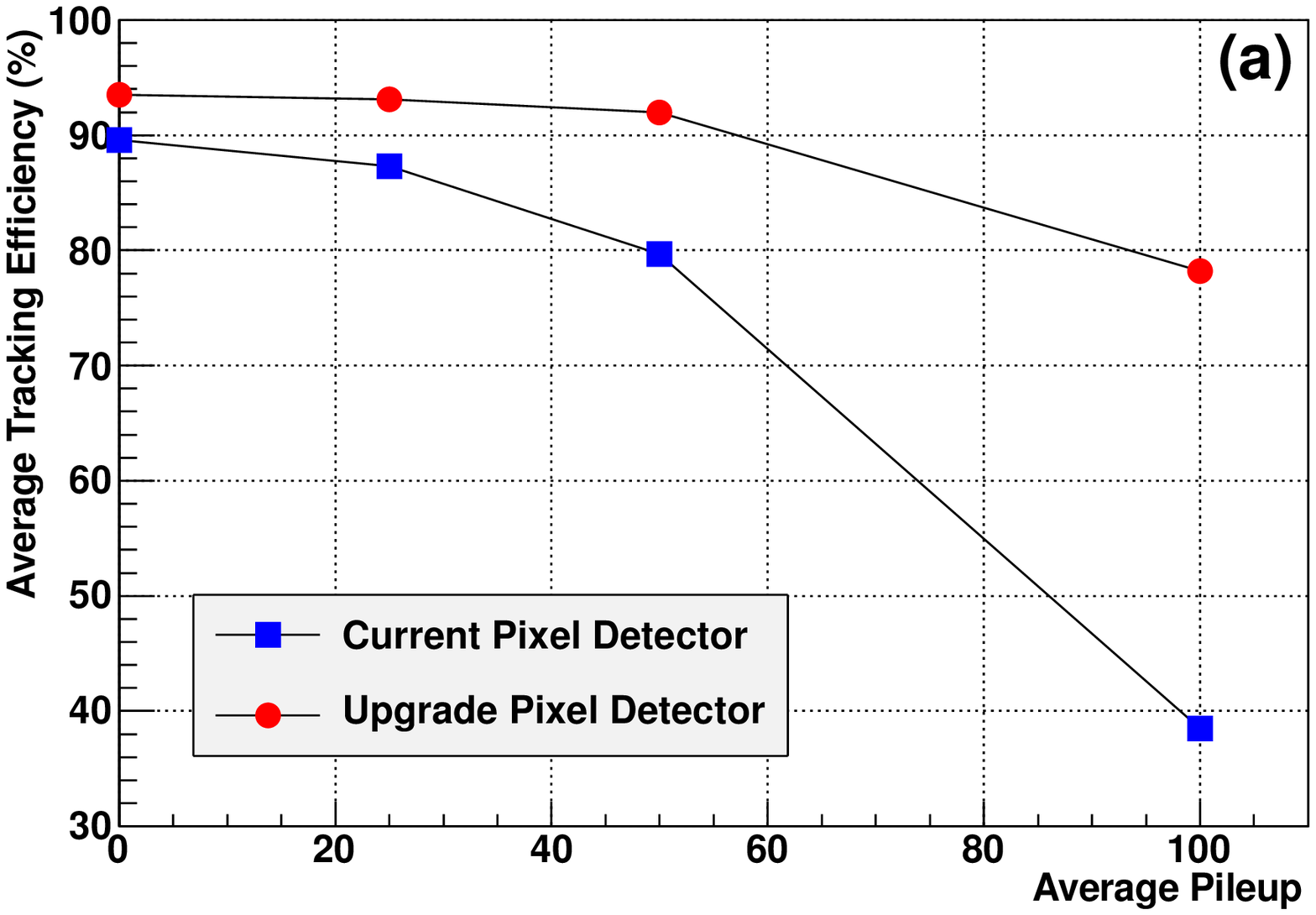}
\hspace*{8pt}
\includegraphics[width=2.5in]{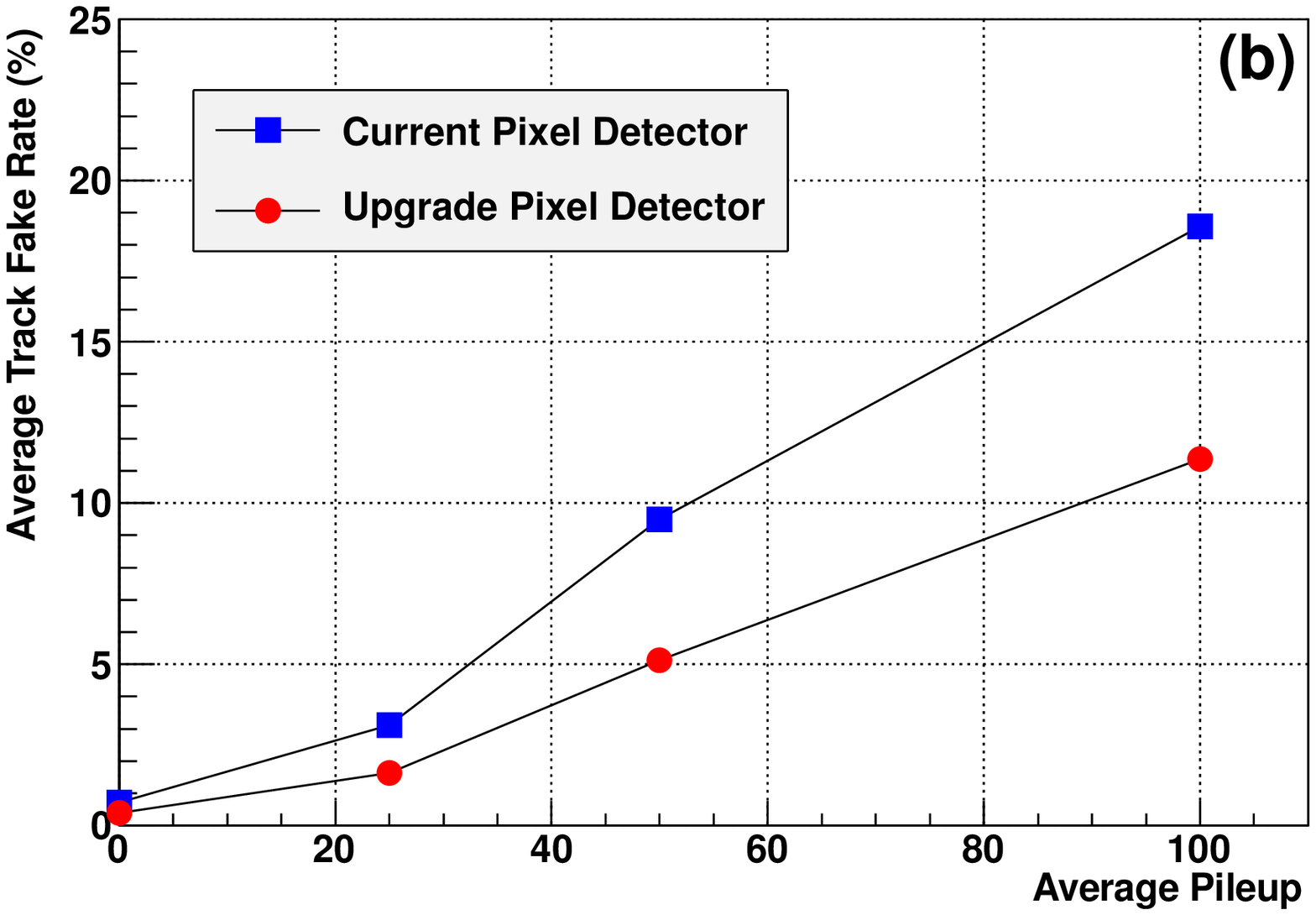}
}
\vspace*{8pt}
\caption{Average tracking efficiency (left) and average track fake rate (right) for the $t \overline{t}$ sample as
a function of the average pileup. Results were determined using the expected ROC data loss
expected for each given average pileup, and for the current pixel detector (blue squares) and
for the upgrade pixel detector (red dots).\protect\label{fig6}}
\end{figure}

The improvement that the new pixel detector brings are higher efficiencies, lower fake rates, lower dead-time and data-loss, and
an extended lifetime of the detector. The expected tracking efficiency and fake rate of the upgraded
 pixel detector for several pileup scenarios in simulated $t \overline{t}$ events are shown in Fig. ~\ref{fig6}.

The Phase 1 detector performance remains excellent at high luminosity. This translates in higher efficiency and lower fake rates for the reconstruction of  $b$-tagged jets which will have an impact on the measurement of $H \rightarrow  b{\overline b}$. It is also impressive the impact that the new detector has on $ H \rightarrow  ZZ $ with $Z \rightarrow  \mu^+ \mu^-$  and $Z \rightarrow  e^+ e^-$. To evaluate the Physics impact on the latter we have used similar techniques to the one develop for the Higgs discovery paper\cite{HiggsDiscovery}  and
 therefore the analysis is optimised  for  a Higgs boson in the mass range $ 110  < m_H < 160$ GeV  produced in proton-proton collision at a center-of-mass energies of 7-8 TeV. 
The analysis has not yet been optimised to the collision environment expect in run 2 where the LHC will provide
a higher center-of-mass energy of 14 TeV and larger pile up. As a benchmark, we have used similar selections as described in \cite{CMS4l} and have
used a Higgs boson MC sample with $m_H = 125$ GeV. The analysis relies on 
the reconstruction and identification of leptons well isolated from other particles in the
event. The Monte Carlo event were generated for the current pixel detector and for the upgrade detector at an average pileup of 50.

\begin{figure}[ph]
\centerline{
\includegraphics[width=4.0in]{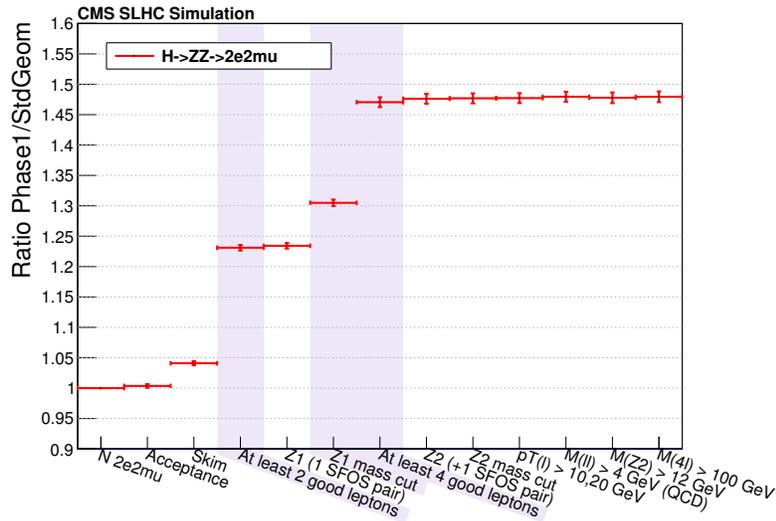}
}
\vspace*{8pt}
\caption{The ratio Phase-1/current of the number of events left after each cut. Values greater
than 1 show increased efficiency for the Phase-1 upgrade and vice versa. The cuts where the
upgraded detector is expected to excel are highlighted.\protect\label{fig7}}
\end{figure}

The Phase-1 pixel detector significantly improves the impact parameter (IP) measurement
of the individual leptons, with respect to the current detector. This improves the
vertex assignment and the vertex reconstruction itself. The former is crucial for this analysis, since 
leptons are required to originate from the primary vertex. This constraint is set by requiring the impact parameter  (IP) to be least than 4 times it error.
The new detector  reduces the RMS on IP by 50\%. In addition the tracking improvements due the Phase 1 pixel detector increase 
the number of leptons passing all the selection requirements. In a large fraction of the events, the number of selected
leptons increases to four. The introduction of the Phase 1 Pixel
detector, with respect to the current detector scenario, lead to an important gain in
signal efficiency. The efficiency ratios, for every step of the analysis, are shown by the cut flow charts in Fig. ~\ref{fig7}  for the  $H \rightarrow 2\mu 2e$  channel.  The efficiency for this channel increases by 48\%. One obtains similar gains for r $H \rightarrow  4\mu$ and $H \rightarrow 4e$ for which we observe improvements in the selection efficiency varying from 41\% to 51\% respectively.

\section{Conclusions}

The discovery of the Higgs boson has opened a new chapter in particle physics. The LHC experiments will continue to study the
properties of the new particle and determine if this particle couples to new physics.
The increase luminosity the LHC during Run 2  brings many challenges. The CMS collaboration 
has designed an upgraded pixel detector that should perform in this environment as well or better
than the current detector does at lower luminosities.
The  improvements that the new detector brings to Higgs physics are substantial and will have a significant impact on the future CMS physics
program. 	

\section*{Acknowledgments}

I would like to thank my colleagues of the CMS experiment and especially the $H\rightarrow ZZ$ sub-group and the team who built the CMS pixel detector and it is now working for its replacement. The author would also like to thank the US DOE and NSF for their support.


\end{document}